\documentclass[apj]{emulateapj}
 
\usepackage{amsmath}

\def\lsim{\mathrel{\lower .85ex\hbox{\rlap{$\sim$}\raise
.95ex\hbox{$<$} }}}
\def\gsim{\mathrel{\lower .80ex\hbox{\rlap{$\sim$}\raise
.90ex\hbox{$>$} }}}
\newbox\grsign \setbox\grsign=\hbox{$>$}
\newdimen\grdimen \grdimen=\ht\grsign
\newbox\laxbox \newbox\gaxbox
\setbox\gaxbox=\hbox{\raise.5ex\hbox{$>$}\llap
     {\lower.5ex\hbox{$\sim$}}}\ht1=\grdimen\dp1=0pt
\setbox\laxbox=\hbox{\raise.5ex\hbox{$<$}\llap
     {\lower.5ex\hbox{$\sim$}}}\ht2=\grdimen\dp2=0pt
\def\gax{\mathrel{\copy\gaxbox}}
\def\lax{\mathrel{\copy\laxbox}}

\shorttitle{GRB 060218/SN~2006aj}
\shortauthors{Mirabal et al.}

\begin{document}

\title{GRB 060218/SN~2006aj: A Gamma-Ray Burst and Prompt Supernova at $z=0.0335$}

\author{N. Mirabal\altaffilmark{1}, J. P. Halpern\altaffilmark{2}, D. An\altaffilmark{3}, J. R. Thorstensen\altaffilmark{4}, \& D. M. Terndrup\altaffilmark{3}}

\altaffiltext{1}{Department of Astronomy, University of Michigan, 
Ann Arbor, MI~48109--1090}
\altaffiltext{2}{Department of Astronomy, Columbia University,
  550 West 120th Street, New York, NY~10027}
\altaffiltext{3}{Department of Astronomy, The Ohio State University, Columbus, 
OH 43210}
\altaffiltext{4}{Department of Physics and Astronomy, Dartmouth College, 
6127 Wilder Laboratory, Hanover, NH 03755-3528}

\begin{abstract}
We report the imaging and spectroscopic localization of GRB 060218 
to a low-metallicity dwarf starburst galaxy at $z = 0.03345 \pm 0.00006$.
In addition to making it the second nearest gamma-ray burst known,
optical spectroscopy reveals the earliest detection of
weak, supernova-like \ion{Si}{2} near 5720~\AA\ ($\sim 0.1 c$), 
starting 1.95 days after
the burst trigger.
$UBVRI$ photometry obtained between 1 and 26 days post-burst
confirms the early rise of supernova light, and suggests a short 
time delay between the gamma-ray burst and the onset of the supernova explosion
if the early appearance of a soft component in the X-ray spectrum
is understood as a ``shock breakout.'' 
Together, these results verify the long-hypothesized origin of soft
gamma-ray bursts in the deaths of massive stars. 
\end{abstract}

\keywords{gamma rays: bursts -- supernovae: 
individual (SN~2006aj) -- supernovae: general}

\section{Introduction}

It is now accepted that 
the so-called long-soft ($\gax$2 s)   
gamma-ray bursts (GRBs) accompany some core-collapse supernovae of 
Type Ic \citep{galama,patat,stanek,hjorth}.
The collective evidence also lends credence to
the collapsar model for GRBs, in which a
relativistic jet breaks through
and explodes a hydrogen-stripped Wolf-Rayet star
\citep{woosley,macf}. Unfortunately,
our understanding of these energetic explosions is still 
limited by the paucity
of nearby ($z \lax 0.2$) GRBs with high-quality photometry and
spectroscopy.

On UT 2006 February 18.149, the {\it Swift\/} Burst Alert Telescope
(BAT) detected an unusually long duration high-energy event \citep{cusumano}.
Its prompt gamma-ray light curve was soft, confined to the 15--50 keV band
for the first 290~s \citep{barbier}; this was followed by a spectrally 
harder (25--100 keV) 10-s ``spike'' that concluded with an exponential
coda extending beyond $t \approx 2000$~s \citep{barthelmy}. 
The relatively odd behavior of this high-energy transient
generated uncertainty as to its basic nature 
\citep[][and references therein]{gehrels},
and poor observing conditions at ground-based
observatories prevented a quick resolution.
However, the identification of an extended optical object
at the precise position of the X-ray and optical transient (OT)
in pre-burst observations of this field \citep{cool,mirabal1}
favored an extra-galactic location.  Ultimately,
optical spectroscopic observations of the OT discovered 
by the {\it Swift\/} UV/Optical Telescope \citep{cusumano} 
confirmed the low-redshift $z = 0.033$, extragalactic nature of this
unusual GRB \citep{mirabal2}, making it the second lowest burst 
redshift known to date after GRB 980425/SN~1998bw at $z=0.0085$. 

In this Letter we describe the identification of the GRB 
060218 host galaxy and its redshift, together with 
photometry and spectroscopy that verify its origin 
in the explosion of a massive star, and discuss the
implications of our results.
We assume an $H_0 = $71 km~s$^{-1}$~Mpc$^{-1}$, $\Omega_m = 0.27$,
$\Omega_{\Lambda} = 0.73$ cosmological model, corresponding
to a luminosity distance $D_{L} = 145$~Mpc at $z = 0.0335$
and an angular scale 0.658 kpc~arcsec$^{-1}$.

\section{Observations}

\begin{figure}[t]
\centerline{
  \includegraphics[width=0.97\linewidth]{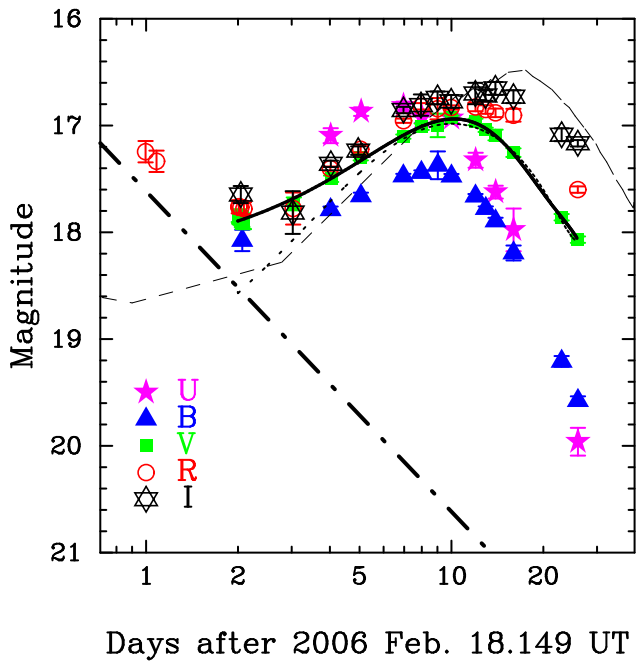}\hfill
}
\caption{$UBVRI$ data for GRB 060218, corrected for Galactic
extinction and host-galaxy contamination.
The {\it solid line} is a fit to the $V$-band light curve.
The {\it dotted line} is a fit to the $V$-band light curve after
subtracting an $\alpha = 1.2$ power-law decay ({\it dot-dashed line})
as justified in the text.
The {\it dashed line} is a template of the $V$-band
light curve of SN~1998bw \citep{galama} shifted to $z=0.0335$.
[{\it See the electronic edition of the Journal for a color
version of this figure.}]}
\label{light_curve}
\end{figure}

Optical observations of GRB 060218 began at the MDM Observatory
on Feb. 19.146 UT using the 2.4m telescope and RETROCAM, the Retractable
Optical Camera for Monitoring,
equipped with SDSS filters \citep{morgan}, and continued on February 20.
Additional $UBVRI$ photometric observations were carried out 
on the MDM 1.3m and 2.4m telescopes using a SITe $2048 \times 2048$ 
thinned, back side-illuminated CCD on several nights from February 21
to March 16.  All of the photometry was converted to a common 
$UBVRI$ system using \citet{landolt} standard stars, and corrected
for Galactic extinction assuming $E(B-V)= 0.142$ from the dust maps of 
\citet{schlegel}.
We note that this value is consistent with the extinction estimated
from high-resolution spectra by \citet{guenther} using
the combined Na~I~D absorption-line equivalent widths from the
Galaxy and GRB host.
This uniform data set, listed in Table~\ref{table1} and shown in 
Fig.~\ref{light_curve}, can be fitted with a power-law decay plus a
supernova (SN) 
light curve that will be described in more detail in the following section. 

The position of the OT in an MDM 2.4m image
was measured with respect to the apparent host galaxy SDSS J032139.68+165201.7
using a set of unsaturated stars common to both images.  We find that the OT
is centered on the compact galaxy to less than $0^{\prime\prime}\!.2$ (130~pc)
in each coordinate. This is to be compared with the galaxy's half-light radius, 
$r_{1/2} \approx  1^{\prime\prime}\!.5$ (1.0~kpc).

\begin{figure}[t]
\centerline{
  \includegraphics[width=0.97\linewidth]{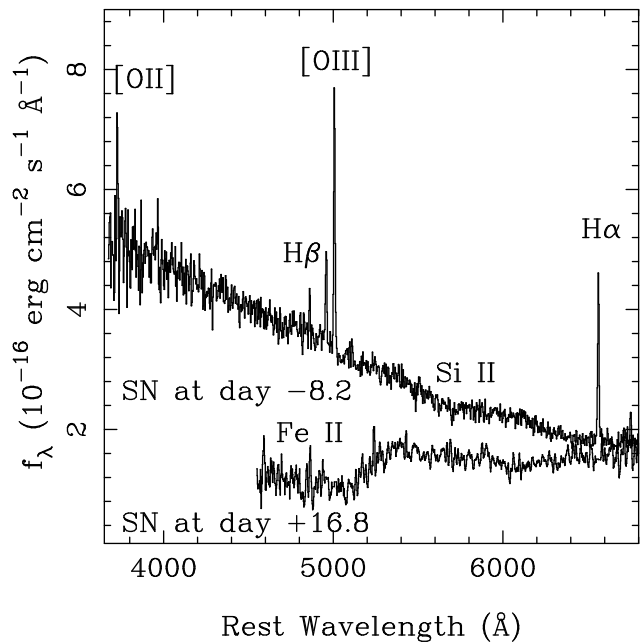}\hfill
}
\caption{Spectra of GRB 060218 obtained on 2006 February 20.097 UT (1.95 days
after the burst) and March 17.12 (27 days post-burst).  Days relative
to supernova maximum are indicated.  Starburst emission lines from the host
galaxy were excised from the second spectrum.
The first spectrum marks the earliest appearance
of Si II near 5720 \AA, while its continuum
is reasonably well fitted by a spectral index $\beta =
0.1 \pm 0.3$ (see text).  Photometry before and after the spectrum was taken
(Table~\ref{table1}) is consistent with this spectral index.}
\label{spectrum}
\end{figure}

Spectra of GRB 060218 were obtained starting on February 20.097 UT
with the Boller \& Chivens CCD spectrograph (CCDS)
mounted on the 2.4 m telescope. The setup used provides 
3.1~\AA\ pixel$^{-1}$ dispersion and $\approx 8.2$~\AA\
resolution with a $1^{\prime\prime}$ slit. The observations consisted of
two 1800~s integrations under fair sky conditions. 
The spectra were processed using standard procedures in IRAF.
\footnote{IRAF is distributed by the National Optical Astronomy Observatories,
    which are operated by the Association of Universities for Research
    in Astronomy, Inc., under cooperative agreement with the National
    Science Foundation.} The wavelength scale was established by fitting 
a set of polynomials to Xe lamp spectra obtained immediately after each target 
exposure.   The spectrophotometric standard star Feige~34 \citep{stone},
observed at comparable telescope pointing to the GRB, was used for
flux calibration.  Although no order-separating filter was used,
we expect that second-order contamination 
is less than 1.5\% below 7000 \AA\ \citep[e.g.,][]{izotov}.  
Another set of spectra, consisting of three 720~s integrations,
was obtained on March 17.12 UT using the Modular Spectrograph (ModSpec) 
on the 2.4 m telescope,
which provides 2~\AA\ pixel$^{-1}$ dispersion and $\approx 3.6$~\AA\ resolution
with a $1.\!^{\prime\prime}1$ slit.  
Similar reduction steps, plus correction for
atmospheric absorption bands, were performed.
Figure~\ref{spectrum} shows the dereddened \citep{cardelli},
wavelength- and flux-calibrated spectra of GRB 060218/SN~2006aj.

\section{Results}

Strong, redshifted nebular emission
lines identified in the spectrum are 
listed in Table~\ref{table2}.  The fluxes
are taken from the CCDS spectrum, while
the more accurate wavelengths from the
ModSpec are listed.
The weighted mean heliocentric redshift derived from the emission lines is 
$z=0.03345 \pm 0.00006$, and the line widths are unresolved at the
resolution of $\approx 160$ km~s$^{-1}$.  The emission-line redshift
is consistent with the Na~I~D absorption-line velocities from the host
\citep{guenther}, falling within the 20~km~s$^{-1}$ spread of the latter.
The Balmer decrement indicates little or no
intrinsic reddening.  
In addition, the earlier spectrum reveals a 
relatively blue continuum that
is reasonably well fitted by a power-law of the form $f
_{\nu} \propto 
\nu^{-\beta}$ with spectral index $\beta = 0.1 \pm 0.3$  
and by a broad P~Cygni feature with the bottom of the 
absorption trough at $\approx 5720$ \AA\ rest wavelength.

We argue that this spectral feature corresponds to 
\ion{Si}{2} $\lambda 6355$ 
with an expansion velocity 31500 $\pm $9200 km~s$^{-1}$. 
An identification with Na~I~D moving at about 8700 km s$^{-1}$,
or with He~I $\lambda 5876$ at even lower velocity,
appears less likely when compared to the early-time optical spectra
of Type Ic SNe \citep{patat}. 
This detection signals the emergence of the supernova designated SN
2006aj \citep{masetti,soderberg,fugazza,
mirabal3,fatkhullin,mazzali,modjaz},
and seals the connection between GRB 060218 and 
the explosion of a massive star.
The weak \ion{Si}{2} line, as well as a nearly featureless
continuum at early times, are typical of Type Ic SN explosions lacking 
both a hydrogen and helium envelope \citep[e.g.][]{filippenko}. SN~2006aj appears to
be much bluer than the Type Ic SN~1998bw.
The early spectral shape of SN~2006aj resembles more closely 
the Type Ic GRB 030329/SN~2003dh spectra obtained within a week 
of that burst \citep{stanek,hjorth}. 

The emergence of a weak, broad Si II feature so
soon after the {\it Swift\/} BAT localization implies that SN~2006aj
began $< 1.95$ days after the GRB 060218. 
This basic picture is confirmed
by the flattening and rise in the optical light curve
on 2006 February 21.181 UT (Fig.~\ref{light_curve}). To better
constrain the SN contribution to the light curve,
we subtracted the host galaxy contribution and, optionally,
a power-law decay assumed to be from relativistic ejecta interacting 
with the circumburst medium.
The host galaxy magnitudes were transformed
from \citet{cool}, \citet{adelman}, and our own Landolt star calibrations.
There is direct evidence 
that emission generated by the relativistic
ejecta began on the first day after the trigger,
when a temporal power-law index $\alpha \approx 1.2$ describes 
the X-ray observations \citep{cusumano2}.
Such decay is slightly steeper than a power-law index  
$\alpha \approx 0.7 \pm 0.3$ that can be fitted to our $R$-band measurements
prior to February 21, but these are consistent because the optical band is 
already affected by the SN rise at this point.

Taking $\alpha \approx 1.2$ as an estimate for the early optical decay rate
({\it the dot-dashed line} in Fig.~\ref{light_curve}),
and subtracting it and the contribution of the host galaxy will produce 
the residual $V$-band SN light curve shown as the {\it dotted line} in Figure
1. Allowing for uncertainties in the
initial optical decay rate, which could be slower than in the X-ray, 
the SN rise is consistent with an origin at the GRB trigger time.
Alternatively, if we ignore the
$R$-band measurements obtained on February 19--20,
and assume that the light curve is completely dominated by 
the SN~2006aj rise after February 21, we get a much flatter SN
light curve (the solid line in Fig.~\ref{light_curve}).
The latter model is less attractive, as it neglects
the first 2 days of bright, decaying optical emission
and points to a supernova time several days before the GRB.

For comparison we show a fit of the
$V$-band light curve of SN~1998bw \citep{galama} shifted to $z = 0.0335$. 
Both the raw and modeled $V$-band light curves
suggest that SN~2006aj reached maximum in $V$
on 2006 February 28.3 UT ($10.2 \pm 0.3$ days after the burst). 
The peak of SN~2006aj occurs earlier than in SN~1998bw, 
and is more like the Type Ic SN~2002ap \citep{galyam,mazzali2}.
The peak absolute magnitude of SN~2006aj 
is $M_{V} = -18.87$. Although it was much bluer than   
SN~1998bw early on, SN~2006aj's maximum is 0.53~mag 
fainter than SN~1998bw, which probably translates into
$\lax 0.5 M_{\odot}$ of ejected $ ^{56}$Ni mass during the explosion 
\citep{iwamoto,woosley4}. 

At a redshift of $z = 0.0335$ the $\gamma$-ray fluence of GRB 060218 corresponds to isotropic energy $E_{\rm iso}$ = $(6.2 \pm 0.3) \times 10^{49}$ ergs,
and the peak luminosity is $L_{\rm p} = (5 \pm 3) \times 10^{46}$ ergs~s$^{-1}$ 
\citep{sakamoto,campana}. 
This energy release is at least an order of magnitude lower than the average 
energy measured in long-duration GRBs and yet a factor of $\sim 20$ 
larger than the intrinsically weak GRB 980425/SN~1998bw \citep{galama}. 
This energy is in fact comparable to the soft X-ray flash XRF 020903 
\citep{sakamoto}. In terms of the X-ray emission, the isotropic X-ray 
luminosity of GRB 060218 at $t = 10$ hr is $L_{X} \sim 10^{43}$ ergs~s$^{-1}$
\citep{cusumano2}.
This is a factor of $10^{3}$ fainter than the sample of GRB X-ray  
luminosities culled by \citet{berger2}. However, this strict comparison 
may not be meaningful, since it does not take
into account the contribution from an earlier flaring period.
In fact, a rough estimate of the X-ray fluence
of GRB 060218 for the first orbit of {\it Swift\/} data (159--2770 s post-trigger)
yields $\approx 2.3 \times 10^{49}$ ergs, a large fraction of the total.

Next, we examine the host galaxy of GRB~060218 in greater detail. The 
observed line ratios are typical of a high-excitation 
starburst galaxy. In particular, the measured H$\alpha$ line flux,
uncorrected for extinction at the host galaxy,
implies a star-formation rate (SFR) equivalent to  
$\approx$ 0.05--0.15 $M_{\odot}$ yr$^{-1}$ \citep{kennicutt}.  
A similar computation of the SFR using [\ion{O}{2}] yields 
$0.09 \pm 0.05$ $M_{\odot}$ yr$^{-1}$.
We also find a good agreement with the SFR derived using an extrapolation of 
the UV continuum luminosity of the host galaxy as tabulated in \citet{cool}.

Turning our attention to host galaxy metallicity, we can determine  
relative abundances from
emission-line strengths of [\ion{O}{2}], H$\beta$, and [\ion{O}{3}] (Table~\ref{table2}) 
by adopting the calibrations in \citet{kewley}. 
Under this approximation, the metallicity is
[O/H] = $-0.34 \pm 0.3$, assuming log(O/H)$_{\odot}$ + 12 = 8.72
\citep{allende}. This value is slightly larger than the 
measurement for the XRF 
031203 host \citep{prochaska}, but it is still among the lowest observed
for GRB hosts. We also estimate
$M_B \approx -16.01$ for SDSS J032139.68+165201.7, 
which ranks this galaxy in 
the bright end of Local Group dwarf galaxies \citep{mateo}. 

\section{Discussion}

Together, its energetics and host-galaxy metallicity place 
GRB 060218 somewhere between 
GRB 980425/SN~1998bw \citep{galama} and XRF 031203 \citep{soderberg3}.
Most importantly, this event continues to shape a picture in which 
sub-luminous, sub-energetic GRB-supernovae born in low-metallicity, dwarf 
galaxies dominate the local ($z \lax 0.5$) population
of GRB events \citep[e.g.][]{soderberg3}. It remains to be seen 
whether this empirical 
result constitutes a selection effect (i.e., we are simply missing the 
analogues of  
faint explosions at higher redshifts) or whether the trend
is indeed a real consequence of stellar and/or metallicity evolution 
over the ages \citep{ramirez1,woosley2,langer}. If the latter is correct, 
low-metallicity progenitors in the local Universe 
may differ from GRB progenitors at higher redshifts.

It is notable that during the early, X-ray bright phase
the absorbing column density needed to fit the {\it Swift\/} X-ray spectrum,
$N_{\rm H} = 6 \times 10^{21}$~cm$^{-2}$ \citep{campana},
is considerably greater than both the Galactic
and host extinction derived from optical emission and absorption lines,
as well as from the optical colors of the afterglow.
The standard Galactic ratio $N$(\ion{H}{1}+H$_2$)/$E(B-V) = 
5.8 \times 10^{21}$ cm$^{-2}$~mag$^{-1}$ \citep{bohlin} would predict
$E(B-V) \approx 1$, rather than the
$E(B-V) \approx 0.17\pm0.03$ observed from the optical methods.
Since ``X-ray $N_{\rm H}$'' is not really $N_{\rm H}$, but a proxy for the 
heavier elements that dominate X-ray photoelectric absorption, this implies
a dust-deficient medium.  The stellar wind of a Wolf-Rayet progenitor
has enough column density to be the location of this excess photoelectric
X-ray absorption.

Phenomenologically, GRB 060218 does share some properties with the ``classical'' 
high-redshift burst population, showing the canonical imprint of emission
from a relativistic blast wave or jet running into circumstellar 
material ($\alpha \approx 1.2$) prior to the SN emergence. It also  
reveals extended prompt X-ray emission remarkably similar to
the early X-ray light curves inferred from {\it Swift\/} X-ray Telescope
observations \citep{zhang2}. Based solely on the early results from
GRB 060218, and using our sparse optical data before the SN rise,
we cannot firmly distinguish among a relativistic jet afterglow,
emission from a jet cocoon \citep{enrico}, or
decaying blackbody radiation associated with ``shock breakout''
\citep{campana}. As calculated in spherical SN models, 
shock breakout refers to the first
observable event after core collapse, and it will occur at $\Delta t
\sim 1~$hr \citep{arnett}.  But in the likely case of a highly asymmetric explosion
accompanying a GRB, a narrow ``jet breakout'' can be observed,
as well as a jet-driven shock emerging with different delays from around
the stellar envelope.

\section{Conclusions}

We reported the identification of GRB 060218 as the second nearest GRB known to date.
Taken together, the emerging SN light curve, the development of a Si II feature, and 
the lack of hydrogen lines in the spectrum indicate that the progenitor of this event
was a massive star, most likely in the Wolf-Rayet family, that was stripped of its hydrogen 
and helium envelope prior to the explosion. 
The presence of early ($\lax$ 2 days after the burst) residual light
in the decay suggests that a fraction of the early emission was created by 
a mildly relativistic blast wave or jet running into circumstellar material. 
Furthermore, the extrapolation of a derived SN~2006aj light curve back in time 
supports the idea that the GRB was nearly simultaneous with the massive core
collapse that gave rise to the SN.
Finally, the sub-energetic nature of this nearby event and its 
location within a low-metallicity host galaxy
highlight the possibility that the variety in massive stellar explosions
is, in part, intrinsic to the metallicity of their progenitors.

\acknowledgments

This work was supported by  
the National Science Foundation under grant 0206051.

\begin{deluxetable*}{lcllcllcc}[]
\tablecolumns{9}
\tablewidth{0pc}
\tablecaption{Optical Photometry of GRB 060218}
\tablehead{\colhead{\hfil Date (UT) \hfil}  & \colhead{Filter} & \colhead{Magnitude\tablenotemark{a}} \hspace{0.3in} &
\colhead{\hfil Date (UT) \hfil}  & \colhead{Filter} & \colhead{Magnitude\tablenotemark{a}} \hspace{0.3in} &
\colhead{\hfil Date (UT) \hfil}  & \colhead{Filter} & \colhead{Magnitude\tablenotemark{a}}}
\startdata
2006 Feb. 22.173 & $U$ & $17.10 \pm 0.07$ & 2006 Feb. 20.208 & $V$ & $17.89 \pm
0.06$ & 2006 Feb. 25.114 & $R$ & $16.96 \pm 0.02$\\
2006 Feb. 23.210 & $U$ & $16.87 \pm 0.04$ & 2006 Feb. 21.196 & $V$ & $17.73 \pm
0.06$ & 2006 Feb. 26.112 & $R$ & $16.82 \pm 0.03$\\
2006 Feb. 25.119 & $U$ & $16.82 \pm 0.04$ & 2006 Feb. 22.186 & $V$ & $17.50 \pm
0.03$ & 2006 Feb. 27.162 & $R$ & $16.84 \pm 0.10$\\
2006 Feb. 26.154 & $U$ & $16.92 \pm 0.05$ & 2006 Feb. 23.196 & $V$ & $17.30 \pm
0.03$ & 2006 Feb. 28.136 & $R$ & $16.83 \pm 0.02$\\
2006 Feb. 28.160 & $U$ & $16.93 \pm 0.04$ & 2006 Feb. 25.112 & $V$ & $17.10 \pm
0.02$ & 2006 Mar. 2.140  & $R$ & $16.83 \pm 0.04$\\
2006 Mar.  2.153 & $U$ & $17.32 \pm 0.07$ & 2006 Feb. 26.140 & $V$ & $17.01 \pm
0.03$ & 2006 Mar. 3.098  & $R$ & $16.85 \pm 0.04$\\
2006 Mar.  4.106 & $U$ & $17.62 \pm 0.06$ & 2006 Feb. 27.167 & $V$ & $17.00 \pm
0.11$ & 2006 Mar. 4.094  & $R$ & $16.88 \pm 0.04$\\
2006 Mar.  6.135 & $U$ & $17.98 \pm 0.20$ & 2006 Feb. 28.168 & $V$ & $16.93 \pm
0.02$ & 2006 Mar. 6.113  & $R$ & $16.90 \pm 0.06$\\
2006 Mar. 16.111 & $U$ & $19.96 \pm 0.13$ & 2006 Mar. 2.143  & $V$ & $16.97 \pm
0.03$ & 2006 Mar. 16.106 & $R$ & $17.60 \pm 0.04$\\
2006 Feb. 20.214 & $B$ & $18.08 \pm 0.10$ & 2006 Mar. 3.102  & $V$ & $17.04 \pm
0.02$ & 2006 Feb. 20.191 & $I$ & $17.65 \pm 0.08$\\
2006 Feb. 22.169 & $B$ & $17.79 \pm 0.03$ & 2006 Mar. 4.097  & $V$ & $17.09 \pm
0.05$ & 2006 Feb. 21.172 & $I$ & $17.82 \pm 0.20$\\
2006 Feb. 23.202 & $B$ & $17.66 \pm 0.03$ & 2006 Mar. 6.118  & $V$ & $17.25 \pm
0.04$ & 2006 Feb. 22.177 & $I$ & $17.36 \pm 0.03$\\
2006 Feb. 25.109 & $B$ & $17.47 \pm 0.02$ & 2006 Mar. 13.112 & $V$ & $17.86 \pm
0.03$ & 2006 Feb. 23.100 & $I$ & $17.24 \pm 0.04$\\
2006 Feb. 26.126 & $B$ & $17.44 \pm 0.02$ & 2006 Mar. 16.108 & $V$ & $18.07 \pm
0.03$ & 2006 Feb. 25.116 & $I$ & $16.86 \pm 0.03$\\
2006 Feb. 27.173 & $B$ & $17.37 \pm 0.13$ & 2006 Feb. 19.146 & $R$ & $17.25 \pm
0.10$ & 2006 Feb. 26.102 & $I$ & $16.81 \pm 0.10$\\
2006 Feb. 28.015 & $B$ & $17.47 \pm 0.02$ & 2006 Feb. 19.230 & $R$ & $17.34 \pm
0.10$ & 2006 Feb. 27.158 & $I$ & $16.74 \pm 0.06$\\
2006 Mar. 2.147  & $B$ & $17.66 \pm 0.02$ & 2006 Feb. 20.162 & $R$ & $17.76 \pm
0.06$ & 2006 Feb. 28.134 & $I$ & $16.77 \pm 0.06$\\
2006 Mar. 3.105  & $B$ & $17.78 \pm 0.02$ & 2006 Feb. 20.168 & $R$ & $17.76 \pm
0.06$ & 2006 Mar. 2.130  & $I$ & $16.70 \pm 0.10$\\
2006 Mar. 4.100  & $B$ & $17.89 \pm 0.03$ & 2006 Feb. 20.191 & $R$ & $17.77 \pm
0.06$ & 2006 Mar. 3.089  & $I$ & $16.72 \pm 0.05$\\
2006 Mar. 6.124  & $B$ & $18.19 \pm 0.07$ & 2006 Feb. 20.240 & $R$ & $17.78 \pm
0.06$ & 2006 Mar. 4.091  & $I$ & $16.65 \pm 0.06$\\
2006 Mar. 13.114 & $B$ & $19.20 \pm 0.05$ & 2006 Feb. 21.180 & $R$ & $17.78 \pm
0.15$ & 2006 Mar. 6.109  & $I$ & $16.73 \pm 0.06$\\
2006 Mar. 16.109 & $B$ & $19.58 \pm 0.04$ & 2006 Feb. 22.161 & $R$ & $17.40 \pm
0.02$ & 2006 Mar. 13.109 & $I$ & $17.09 \pm 0.06$\\
2006 Feb. 20.162 & $V$ & $17.84 \pm 0.06$ & 2006 Feb. 23.192 & $R$ & $17.22 \pm
0.03$ & 2006 Mar. 16.105 & $I$ & $17.17 \pm 0.04$\\
\enddata
\tablenotetext{a}{Host galaxy with
$U = 20.10$, $B = 20.41$, $V = 20.09$,
$R = 19.91$, $I = 19.54$ was subtracted,
then remainder was corrected for Galactic extinction
$A_{\rm U} = 0.77$, $A_{\rm B} = 0.61$, $A_{\rm V} = 0.47$,
$A_{\rm R} = 0.38$, and $A_{\rm I} = 0.28$ respectively.}
\label{table1}
\end{deluxetable*}
 
\begin{deluxetable}{lcc}[]
\tablecolumns{3}
\tablewidth{0pc}
\tablecaption{Host-Galaxy Emission Lines}
\tablehead{
\colhead{Line $\lambda_{\rm air}$(\AA)} & \colhead{$\lambda_{\rm helio}$(\AA)} & \colhead{$F(\lambda)/F({\rm H}\beta$)\tablenotemark{a}}
}
\startdata
[\ion{O}{2}] 3727.5    & (3850.7)\tablenotemark{b} & $2.61 \pm 0.4$ \\
 
H$\beta$ 4861.33       &  5024.36   & $1.00 \pm 0.3$ \\
 
[\ion{O}{3}] 4958.92   &  5124.84   & $2.03 \pm 0.3$ \\
 
[\ion{O}{3}] 5006.85   &  5174.31   & $5.09 \pm 0.6$ \\
 
H$\alpha$ 6562.79      &  6782.25   & $2.94 \pm 0.3$ \\
 
[\ion{N}{2}] 6583.39   &  . . .     & $<0.08$ \\
 
[\ion{S}{2}] 6716.42   &  . . .     & $<0.65$ \\
 
[\ion{S}{2}] 6730.78   &  . . .     & $<0.54$ \\
 
\enddata
\tablenotetext{a}{Flux relative to $F({\rm H}\beta) = 9.94\times10^{-16}$
ergs~cm$^{-2}$~s$^{-1}$, corrected for Galactic extinction $E(B-V) = 0.142$.}
\tablenotetext{b}{Poor wavelength calibration in this region.}
\label{table2}
\end{deluxetable}

\end{document}